\documentstyle[11pt,aaspp4,psfig]{article}   

\begin{document}

\def\Msun{{\rm M}_\odot}
\def\msun{{\rm M}_\odot}
\def\rsun{{\rm R}_\odot}
\def\lsun{{\rm L}_\odot}
\def\half{{1\over2}}
\def\RL{R_{\rm L}}
\def\zs{\zeta_{s}}
\def\zR{\zeta_{\rm R}}
\def\dJJ{{\dot J\over J}}
\def\dMM{{\dot M_2\over M_2}}
\def\tKH{t_{\rm KH}}
\def\dd{{\rm d}}
\def\be{\begin{equation}}
\def\ee{\end{equation}}
\def\dmc{\dot M_{\rm crit}}
\def\dmirr{\dot M_{\rm irr}}
\def\ace{\alpha_{\rm CE}}
\def\rf{r(f>f_0)}
\def\vmean{{\langle V_k^2 \rangle}^{1/2}}

\slugcomment{To appear in the Astrophysical Journal}

\title{Supernova Kicks, Magnetic Braking, and Neutron--Star Binaries} 
\author{V.~Kalogera\altaffilmark{1,2}, U.~Kolb\altaffilmark{3,4}, and 
A.\,R.~King\altaffilmark{3}}

\altaffiltext{1}{Harvard-Smithsonian Center for Astrophysics, 60 
Garden St., Cambridge, MA 02138; vkalogera@cfa.harvard.edu} 
\altaffiltext{2}{Astronomy Department, University of Illinois, Urbana, IL 
61801} 
\altaffiltext{3}{Astronomy Group, University of Leicester, 
Leicester LE1 7RH, U.\,K.; uck@star.le.ac.uk, ark@star.le.ac.uk}
\altaffiltext{4}{Max--Planck--Institut f\"ur Astrophysik,
Karl--Schwarzschild-Str.~1, 85740~Garching, Germany}

\begin{abstract} 
We consider the formation of low--mass X--ray binaries containing
accreting neutron stars via the helium--star supernova channel.  The
predicted relative number of short--period transients provides a
sensitive test of the input physics in this process.  We investigate the
effect of varying mean kick velocities, orbital angular momentum loss
efficiencies, and common envelope ejection efficiencies on the
subpopulation of short--period systems, both transient and persistent. 
Guided by the thermal--viscous disk instability model in
irradiation--dominated disks, we posit that short--period transients
have donors close to the end of core--hydrogen burning. We find that
with increasing mean kick velocity the overall short-period fraction,
$s$, grows, while the fraction, $r$, of systems with evolved donors
among short-period systems drops.  This effect, acting in opposite
directions on these two fractions, allows us to constrain models of LMXB
formation through comparison with observational estimates of $s$ and
$r$.  Without fine tuning or extreme assumptions about evolutionary
parameters, consistency between models and current observations is
achieved for a regime of intermediate average kick magnitudes of about
100--200\,km\,s$^{-1}$, provided that (i)~orbital braking for systems
with donor masses in the range $1-1.5\,\msun$ is weak, i.e., much less
effective than a simple extrapolation of standard magnetic braking
beyond $1.0\,\msun$ would suggest, and (ii)~the efficiency of common
envelope ejection is low. 
\end{abstract}


\section{INTRODUCTION}

The structure and properties of accretion disks around non-magnetic compact
objects in close binaries are primarily determined by the rate at which
matter is supplied by the Roche--lobe filling donor star. If this rate is
smaller than a critical value $\dmc$, the disk is subject to a
thermal--viscous instability and undergoes a limit cycle evolution
alternating between a hot and a cool state, the outburst and quiescent
phases. This model has been successfully applied to dwarf nova outbursts in
cataclysmic variables (CVs), where the accretor is a white dwarf (WD) (for
reviews see, e.g., Cannizzo\markcite{C93} 1993; Osaki\markcite{O96} 1996),
and also to X--ray transient outbursts in low-mass X-ray binaries (LMXBs),
where the accretor is a neutron star or a black hole (for reviews
see, e.g., Lasota\markcite{L96} 1996; Wheeler\markcite{W98} 1998). 

In the case of LMXBs, where the accretion efficiency and hence the X-ray
luminosity from the compact object is higher than in CVs, there is strong
evidence that self--irradiation is important (see, e.g., van~Paradijs \&
McClintock\markcite{vP94} 1994; King \& Ritter\markcite{KR98} 1998).  X--ray
irradiation from the central accretor raises the disk temperature for a
given mass transfer rate, thereby suppressing the instability.  As a
consequence, $\dmc$ is significantly smaller in LMXBs than in CVs,
consistent with observations (van~Paradijs\markcite{vP96} 1996). The
strength of the irradiation in LMXBs depends on the nature of the compact
object. In the neutron--star case, the irradiating source is equivalent to
a point source at the center of the disk, while, in the black--hole case,
the lack of a hard stellar surface implies that the irradiating source is
only the innermost disk, which is weaker by a factor about equal to the
relative disk thickness, the other system parameters being constant
(King, Kolb, \& Szuskiewicz\markcite{KK97} 1997). 

Disk irradiation has important consequences for short--period LMXBs. 
These systems evolve towards shorter orbital periods $P$ under the
influence of orbital angular momentum losses $\dot J$ (henceforth
``j--driven'' systems), caused by a magnetic stellar wind from the donor
star (magnetic braking) or by gravitational radiation. Assuming that
$\dot{J}_{\rm MB}$ from magnetic braking (MB) is independent of the
nature of the accretor, the same orbital braking formalism must apply
for both CVs and LMXBs.  For a given MB law, i.e., for a given
dependence of $\dot J$ on binary parameters, the mass transfer rate
(proportional to the fractional angular momentum loss rate) is smaller
in black--hole binaries than in neutron--star binaries, as the total
angular momentum increases with primary mass. This, together with the
higher values for $\dmc$, leads to transient behavior for all j--driven
black--hole binaries, as observed. The converse appears to be true for
neutron--star systems. The MB transfer rate may be up to two orders of
magnitude larger than $\dmc$ (King, Kolb, \& Burderi\markcite{KK96}
1996), suggesting that there are no or only very few neutron star
transients.  However, it is clear from observations that the fraction of
transients among short--period neutron--star LMXBs is non--negligible.
Five out of 23 LMXBs with a confirmed or possible neutron star primary
and with periods in the range $3\,{\rm h} < P < 20\,{\rm h}$ are
classified as transient (Ritter \& Kolb\markcite{R98} 1998).  Although
it is difficult to estimate the {\em intrinsic} value of this fraction,
it may be even larger than $5/23$ (see \S\,5.1 below).

A possible resolution of this apparent contradiction between the
disk-instability model and the observations has been suggested by King,
Kolb, \& Burderi\markcite{KK96} (1996) and King \& Kolb\markcite{K97}
(1997, hereafter KK97). The mass transfer rates in j--driven
LMXBs\footnote{Hereafter, we will refer to neutron--star LMXBs simply as
LMXBs, unless otherwise stated.} with somewhat evolved donor stars (close
to the end of core--hydrogen burning), are significantly smaller than in
systems with unevolved donors, therefore favoring transient behavior. Then
the large observed transient fraction demands that the contribution of
these evolved systems to the total population is correspondingly large. 
Under the assumption of spherically symmetric supernovae (SN) and for
strong magnetic braking at high ($\gtrsim 1.2\,{\rm M}_\odot$) donor star
masses, KK97 showed that LMXBs forming via the standard evolutionary
channel involving a helium star supernova (Sutantyo\markcite{S75} 1975; 
van den Heuvel\markcite{vdH83} 1983) do indeed have this property. However,
for weaker magnetic braking (constrained by rotational velocity data for F
stars), Kalogera \& Webbink\markcite{K98} (1998; hereafter KW98) showed
that j--driven LMXBs form only if supernovae are asymmetric. KK97 also
showed, in a qualitative way, that asymmetric SNe with on average large
kick velocities imparted to the neutron stars would inevitably lead to a
large number of unevolved systems in the population, and hence a small
number of transient systems among the short-period LMXB population,
contrary to observations. 

This qualitative conclusion by KK97 appears to be in contradiction to
numerous pieces of evidence for the existence of rather substantial kick
velocities (Kaspi et al.\ \markcite{Kp96}1996; Hansen \&
Phinney\markcite{H97} 1997; Lorimer, Bailes, \& Harrison\markcite{L97}
1997; Fryer \& Kalogera\markcite{F97} 1997; Fryer, Burrows, \&
Benz\markcite{F98} 1998), so the need for a quantitative study of the
problem arises. The formation of LMXBs including the effect of natal
neutron--star kicks has been studied in detail by KW98.  Here, we use
these detailed synthesis models to address the issue of the transient
fraction among LMXBs and investigate its dependence on different
evolutionary parameters. The increase of this fraction in the model
populations is accompanied by an increase of the fraction of systems
with donors that have evolved beyond the main sequence, i.e., of
long--period systems driven by nuclear expansion of the secondary
(``n--driven'' systems).  Our goal is twofold: first, to disentangle the
differential dependences on model parameters of the two observable
quantities, the predicted fraction of ``j--driven'' systems among all
LMXBs and the predicted transient fraction among ``j--driven'' LMXBs.
Second, to constrain quantitatively the input parameters by requiring
that both predicted fractions are consistent with current observations. 
These two observational constraints operate in opposite directions, and
this allows us to derive limits on the mean magnitude of natal kicks
imparted to neutron stars, the strength of magnetic braking, and the
efficiency with which orbital energy is consumed during the common
envelope phase prior to the supernova.  As the observational sample
increases in the future, the model calculations presented here can be
used to tighten these constraints still further. 

In \S\,2, we review the evolutionary picture of j--driven LMXBs and
derive a simplified criterion for transient behavior. In \S\,3, we describe
the different models for LMXB formation considered here, while the results
and basic effects of varying model parameters are presented in \S\,4. In
\S\,5, we first (\S\,5.1) describe the observed sample and derive the
observational constraints, and then (\S\,5.2) evaluate the models based
on these constraints. We conclude in \S\,6 with a discussion of our results
and of possible extensions of the present study.

\section{EVOLUTION AND TRANSIENT BEHAVIOR}

In the context of the thermal--viscous instability model, transient behavior
occurs when the average mass transfer rate from the donor to the disk is
lower than a critical rate, $\dmc$, so that hydrogen is not ionized at the
outer disk rim.  
As shown in King, Kolb, \& Szuszkiewicz (1997) under reasonable
assumptions for various disk properties, the critical transfer rate for
LMXBs can be approximated by
\be
 \dmc \simeq  6 \times 10^{-12} \, m_{\rm NS}^{5/6} \, m_d^{-1/6} \,
              P_h^{4/3} \, \msun {\rm yr}^{-1}  
\label{3.0}
\ee
(see their eq.\,[7], scaled appropriately for irradiation in neutron star
systems), where $m_{\rm NS}$ and $m_{\rm d}$ are the neutron and
donor star masses, $M_{\rm NS}$ and $M_d$, in solar units, and $P_h$ the
orbital period in hours.  

In general, the secular mean mass--transfer rate 
is determined by the rate of angular momentum loss from the system,
the donor's nuclear expansion rate, and the radius--mass exponents,
$\zeta$, i.e., the logarithmic derivatives of the donor and
Roche--lobe radii with respect to mass (see, e.g., Webbink\markcite{W85} 
1985, 
Ritter\markcite{R96} 1996). For j--driven systems, angular momentum losses 
dominate 
over nuclear evolution. The mean mass transfer rate for such systems is
\be
  \frac{\dot M_d}{M_d} = \frac{2}{\zeta_{\rm eq} - \zeta_{\rm L}}
                            \frac{\dot J}{J} ,
\label{3.2}
\ee
(e.g., Ritter 1996), where the thermal equilibrium radius--mass exponent is
$\zeta_{\rm eq} \sim 1$, the Roche--lobe exponent is $\zeta_{\rm L}\simeq
2M_d/M_{\rm NS} - 5/3$ (for conservative mass transfer), and $J$ is the
orbital angular momentum. The angular momentum loss rate, $\dot J$, is the
sum of the loss rate due to magnetic braking, $\dot J_{\rm MB}$, and due to
gravitational wave radiation, $\dot J_{\rm GR}$. Typically, magnetic braking
is more efficient than gravitational radiation by about two orders of
magnitude at binary periods of a few hours. 

Neither the strength of magnetic braking nor its functional dependence on
stellar and binary parameters is known from first principles.  There are as
many different suggested MB laws as attempts to model the
magnetohydrodynamic problem (e.g., Verbunt\markcite{V84} 1984; Mestel \& 
Spruit\markcite{M87} 1987;
Kawaler\markcite{K88} 1988; Tout \& Pringle\markcite{T92} 1992; Zangrilli, 
Tout, \& Bianchini\markcite{Z97} 1997). For
this reason, an empirical law (Verbunt \& Zwaan\markcite{V81} 1981) derived 
from rotational velocity data of isolated G stars as a function of time
(Skumanich\markcite{S72} 1972) has usually been used. Application of 
magnetic braking to the evolution of close binaries, however, requires the
extrapolation of this empirical law to stars with a different mass and a much
more rapid rotation. 

Adopting a Skumanich-type parameterization of the magnetic braking law, 
$\dot{J}_{\rm MB}~\propto~\Theta\,\omega^3$, where $\omega$ is the orbital 
angular velocity and $\Theta$ the donor's moment of inertia, the 
fractional angular momentum loss rate can be expressed as:
\be
\frac{\dot{J}_{\rm MB}}{J}~=~- \gamma_{\rm 0} \, b(M_d) \, M_d^{\mu} \,
     R_d^{\rho} \, \frac{(M_{\rm NS}+M_d)^{1/3}}{M_{\rm NS}} \,
     P^{-10/3}, 
\ee
where $\gamma_{\rm 0}$ is a normalization constant, 
$b(M_d)$ is a mass-dependent efficiency factor, and
$R_d$ is the radius of the secondary.  The free parameters $\mu$ and
$\rho$, which are believed to be of order unity, allow for a different 
overall mass-- and
radius--dependence. Assuming that the empirical MB law is universal, i.e.,
it is applicable to CVs and LMXBs alike, and that the CV period gap is
caused by a discontinuous drop of $\dot J_{\rm MB}$ to zero at the upper
edge of the gap, it is then possible to calibrate the strength of magnetic
braking, i.e., fix $\gamma_{\rm 0}$ at $P_h\simeq 3$, so that the width and
location of the observed period gap is reproduced in the models (e.g.,
Kolb\markcite{KU96}
1996; Kolb \& Baraffe\markcite{KB98} 1998).  In stars more massive than 
$\simeq 1\,{\rm
M}_\odot$, the outer convective envelope, thought to sustain the stellar
magnetic field, gradually disappears and magnetic braking must become less
efficient. For the extrapolation of the MB law to these higher masses,
one can derive a mass--dependent efficiency, $b(M_d)$,
to reproduce the mean 
rotational velocities of stars of spectral type F0 and F5 (KW98). 

Given the above normalizations, the strength of magnetic braking at
periods longer than a few hours is determined by the dependences on
stellar radius and orbital period. For a binary with a given secondary
mass, we can show that the magnetic braking strength decreases with
increasing degree of nuclear evolution of the donor at the onset of mass
transfer. Using equations (1), (2), and (3), and the fact that in a
semi--detached binary the orbital period is inversely proportional to
the square root of the donor's mean density, i.e., $P \propto R_d^{3/2}$
at a fixed $M_d$, we obtain
\be
\frac{\dot{M}_d}{\dmc} \propto \frac{R_d^{\rho-5}}{R_d^2} \propto 
R_d^{\rho-7}
\label{mdotratio} 
\ee
for the ratio of the mean transfer rate from the donor to the critical
rate for disk instability. If $\rho < 7$, which holds for MB laws
derived both empirically and from first-principle calculations (e.g.,
Verbunt \& Zwaan 1981; Rappaport, Verbunt, \& Joss\markcite{V83} 1983;
Mestel \& Spruit\markcite{M87} 1987), then at fixed donor mass the ratio
$\dot{M}_d/\dmc$ decreases with increasing stellar radius, hence with
orbital period at Roche--lobe overflow. It is therefore lower for donors
that have evolved significantly away from the zero--age main sequence. 
These systems are thus prone to transient behavior. 

A measure of the degree of nuclear evolution of a star is the ratio
$f\equiv t/t_{\rm MS}$, i.e., the age of the donor in units of its
main--sequence lifetime, $t_{\rm MS} \simeq 10^{10} {\rm yr}~m_2^{-3}$.
Secondaries of j--driven systems are still on the main sequence and hence
correspond to $f<1$.  From relation (\ref{mdotratio}) it follows
that transient behavior in a binary can occur if the value of $f$ exceeds
some critical value $f_{\rm 0}$ close to unity. An accurate estimate of
$f_{\rm 0}$ requires detailed models of the evolution of systems with
moderately evolved secondaries of different initial masses and for
different MB laws. Such sequences are computationally demanding and beyond
the scope of this study. The few such computations that have been published
(Pylyser \& Savonije\markcite{P88}\markcite{P89} 1988, 1989; Singer, Kolb,
\& Ritter\markcite{S93} 1993; see also Ritter\markcite{R94} 1994) indicate
that only stars with $f \ga 0.8-0.9$, i.e., close to the end of 
core--hydrogen burning, lead to transfer rates lower than $\dmc$. 
Therefore, in
our discussion of the models, we avoid quantifying $f_0$ and treat it as a
free parameter. However, when comparing to observations, we assume for
simplicity that there is a global value $f_0$ characterizing the transient
population (i.e.\ we neglect its mass dependence and assume that during the
secular evolution $f_0$ does not change), $f_{\rm 0} \simeq 0.8$.

\section{MODELS OF LMXB FORMATION}

Evolution of binaries with initially extreme mass ratios through a
common-envelope phase and the supernova explosion of a helium star is
thought to be the main evolutionary channel leading to LMXB formation. A
detailed quantitative study of this channel, taking into account the
complete set of structural and evolutionary constraints imposed on the
LMXB progenitors as well as the effect of natal neutron star kicks, has
been presented by KW98 (see also Terman, Taam, \& Savage\markcite{T96}
1996). Results of these calculations include predicted model
distributions of nascent (at the onset of mass transfer) LMXBs over
donor mass and orbital period. Here, we extend these calculations for a
set of different magnetic braking laws and use the resulting LMXB
distributions to calculate two quantities that quantitatively describe
the j-driven LMXB population and the transient systems among them. For a
description of the input distributions to the synthesis models, the
relevant constraints on the population, the semi-analytic synthesis
method used, and the dependence of some of the results on the model
parameters, we refer to the extensive discussion in KW98. 

For our purposes here we focus on the following two diagnostic
quantities:  

(i) the fraction, $s$, of j--driven systems among the total neutron--star LMXB
population. Detailed calculations (Pylyser \& Savonije 1988, 1989) of the
secular evolution of a few systems in the mass range of interest here have
shown that j--driven and n--driven systems are separated by a critical
``bifurcation'' period, $P_{\rm crit}$, which in detail depends on the
donor mass and on the magnetic braking strength. Guided by the results of
Pylyser \& Savonije (1988, 1989), we adopt a mass-independent critical
period of $20$\,h (see also \S\,5.1). An alternative definition of $s$, as
the fraction of systems with donors on the main sequence at the onset of
mass transfer, gives essentially the same quantitative results (within
just a few per cent). 

(ii) the normalized cumulative distribution over $f$ of systems with donors
on the main sequence and masses below $\simeq 2$\,M$_\odot$, i.e., the
fraction $r(f>f_0)$ of j--driven systems with donors that are more evolved
than $f_0$, as a function of $f_0$. As is clear from the discussion in
\S\,2, we identify this fraction, for an appropriate value of $f_0$, with
the transient fraction among short--period LMXBs. The upper limit for the
donor mass arises because we require that the system is stable against
thermal--timescale mass transfer.  In the case of conservative mass
transfer, this limit is $1.5-1.6\,\msun$ (with $M_{\rm NS} = 1.4\,\msun$),
while in the case of Eddington--limited mass transfer (Kalogera \&
Webbink\markcite{KW96} 1996) it is $\simeq 2\msun$. 

In the following sections, we study
the dependence of the above two quantities on
the various unknown model parameters. Model assumptions regarding the birth
rate of primordial binaries with extreme mass ratios (of relevance to LMXB
formation) have been shown (KW98) to affect essentially only the predicted
{\em absolute} birth rate of the LMXB population and not its distribution over
binary parameters. Since both of the above diagnostic quantities are ratios
of birth rates, they are not affected by differences in the absolute
normalization.

For the treatment of the common-envelope phase (for a review see Iben \&
Livio\markcite{I93} 1993), we adopt the simplified parametrization of
orbital shrinkage based on the common--envelope efficiency, $\alpha_{\rm
CE}$.  This is the efficiency with which orbital energy is deposited in
the common envelope, causing it to become gravitationally unbound (for
the exact definition used here, see KW98). We investigate models with
high ($\ace=1.0$) and low ($\ace=0.3$) efficiency. 

A major event in the evolution of LMXB progenitors is the explosion of the
primary as a supernova. Studies of a wide variety of neutron star systems,
single and binary radio pulsars, and X--ray binaries, have provided strong
evidence that kicks, probably associated with some asymmetry in the
explosion, are imparted to neutron stars at birth.  However, the
distribution of these kicks in magnitude and direction is not yet known; we
assume it is Gaussian in each of the three velocity components, which leads
to isotropic kicks with a Maxwellian distribution in magnitude. We
investigate models with r.m.s.\ velocities $\langle V_k^2 \rangle ^{1/2} =
20$, $50$, $100$ and $300$\,km\,s$^{-1}$. 

Finally, we consider three different formulations for magnetic braking,  
which we label MB1, MB2 and MB3.

MB1 (``strong'' MB) follows the description by Verbunt \& Zwaan 
(1981), has been used by KK97, and is given by: 
\be
\frac{\dot{J}_{\rm MB}}{J}~=~- 1.8\times 10^{-5}~\frac{(m_{\rm 
NS}+m_d)^{1/3}}{m_{\rm NS}}~r_d^{4}~P_h^{-10/3}~{\rm yr}^{-1},
\ee
where $r_d$ is the donor's radius in solar units.

MB2 (``weak'' MB) 
follows Rappaport, Verbunt, \& Joss (1983) and is given by: 
\be
\frac{\dot{J}_{\rm MB}}{J}~=~- 6.9\times 10^{-6}~\frac{(m_{\rm
NS}+m_d)^{1/3}}{m_{\rm NS}}~r_d^{2}~P_h^{-10/3}~{\rm yr}^{-1}.
\ee

Both expressions are widely used in studies of CV and LMXB evolution. They
are in reasonable agreement with the normalization that reproduces the CV
period gap.  For the extrapolation of the above laws to masses higher than
$\simeq 1$\,M$_\odot$, we consider two variations of them, labeled as a and
b. We adopt either a discontinuous drop to zero at $m_d=1.5$ (laws
MB1a and MB2a), 
or a gradual decline for $m_d>1.03$ with $b(m_d) = \exp
\{-7.23(m_d-1.03)\}$ (law MB1b) and $b(m_d) = \exp \{-4.15(m_d-1.03)\}$
(law MB2b). The latter law was adopted by KW98, and details about the
derivation of $b(M_d)$ are described in that paper. 

We also investigate a third law, MB3, which simply assumes that magnetic 
braking  is
restricted to semi--detached binaries only. In this case, unstable disk
accretion still requires the donor to be close to the end of core--hydrogen
burning, while in the {\em detached} phase prior to LMXB formation and
after the SN explosion, angular momentum is lost only via gravitational
radiation, at a rate
\be
\frac{\dot{J}_{\rm GR}}{J}~=~- 1.25\times 10^{-8}~\frac{m_{\rm NS}}{(m_{\rm
NS}+m_d)^{1/3}}~m_d~P_h^{-8/3}~{\rm yr}^{-1}.
\ee
(e.g., Landau \& Lifshitz\markcite{L51} 1951). The physical motivation for 
this choice is
that the donors in detached binaries need not necessarily be in corotation. 
Then magnetic braking just brakes the donor's spin, not the orbital 
motion. Clearly, MB3
is a somewhat artificial and extreme case included here only to show the
effect of negligible magnetic braking on the two diagnostic quantities, 
$s$ and $r(f>f_{\rm 0})$. 

In Figure \ref{MBlaws}, we plot $\dot J/J$ for all five cases as a function
of donor mass for donors on the zero-age main sequence ($m_{\rm NS}=1.4$;
$P_h=12$).  In the regime $m_d\ga 1$, MB1a is much stronger than MB2a
because of the stronger dependence on $R_d$. In contrast, the laws MB1b and
MB2b are similar for $m_d>1.03$ as the braking efficiency factors $b(m_d)$
were obtained to fit the same rotational velocities of F stars.

\section{RESULTS}

For a set of different models, where the r.m.s.\ kick magnitude, the common
envelope efficiency, and the magnetic braking law were varied, we calculate
the two diagnostic quantities, $s$ and $r(f>f_{\rm 0})$ (see \S\,3), for
the nascent LMXB population. The dependence of the two quantities on the
varying model parameters can be quite strong; in this section we discuss
the various trends as we vary each of the parameters separately and try to
understand the origin of the effects. 
 
{\em Average kick magnitude.} Natal neutron star kicks affect the LMXB
population primarily in two ways (Hills\markcite{H83} 1983; Brandt \&
Podsiadlowski\markcite{B95} 1995; Kalogera\markcite{K96} 1996): (i) systems
that would have been disrupted in their absence remain bound, and (ii)
post-SN orbital separations smaller than the pre-SN ones are allowed and
can actually be more abundant in the post-SN separation distribution. Both
effects favor the formation of short-period systems with unevolved donors.
Compared to the case of zero kicks, the first effect allows LMXBs with
lower donor masses to form (see Fig.\ \ref{mdist}), i.e.\ reduces the mean
donor mass and lengthens the mean nuclear-evolution time, $t_{\rm MS}$. The
second effect reduces the average lifetime of the detached phase as the
magnetic braking strength is inversely proportional to a high power of the
orbital period. Hence the mean value of $f$ is smaller in the presence of
kicks.  

As the average kick magnitude increases, the fraction of short-period
systems increases, while the fraction of j--driven systems with evolved
donors decreases (see Fig.\,\ref{r1}, \ref{r2}, and \ref{s}). In fact, as
$\langle V_k^2 \rangle ^{1/2}$ changes from being negligible to moderate,
the increase of $s$ is extremely steep for {\em all} models, except for the
one with strong magnetic braking (MB1a) and low common-envelope efficiency
($\ace =0.3$).  The origin of this strong dependence is related to the size
of the pre--SN orbits. The combination of all the constraints imposed on the
progenitors leads to such pre--SN orbital sizes that, in the absence of
kicks, only wide (n--driven) LMXBs can be formed, while in the presence of
even small kicks, systems with smaller orbits are allowed to form and the
short-period fraction increases rapidly (KW98). The exception in the case
of MB1a and low values of $\ace$ arises because this particular MB law is
strong enough at high masses to bring a significant fraction of the
population to shorter periods (see KK97). 

{\em Common--envelope efficiency.} A decrease in the value of $\ace$ results
in smaller post--CE orbital separations, therefore smaller post--SN orbits
and an increase in the fraction, $s$, of short-period systems (see
Fig.\,\ref{s}). This effect on $s$ becomes more prominent at low kick
magnitudes, where the vast majority of LMXBs form with wide orbits {\em
unless} $\ace$ is low.  The effect of a decreasing $\ace$ on the
transient fraction among j--driven systems is more complicated and depends
on the strength of magnetic braking, as well.  In the case of strong
magnetic braking (MB1a), the shift of the population to shorter periods
decreases the transient fraction (fraction of systems with high values of
$f$), while in cases of weak or moderate magnetic braking the accompanying
increase of the mean donor mass dominates, increasing the transient
fraction accordingly (see Fig.\,\ref{r1}). 

{\em Magnetic Braking.} As already mentioned, strong magnetic braking
favors the formation of short--period (j--driven) LMXBs. Therefore, the
extreme case of magnetic braking being non-operative during the
semi-detached phase (law MB3) always leads to a LMXB population with the
highest fraction of evolved donors among j--driven systems but at the same
time the lowest short-period fraction in the total population (Fig.\
\ref{r1} and \ref{s}). The effect of varying the magnetic braking strength
is also more prominent when the other two parameters, average kick
magnitude and common-envelope efficiency, are such that the majority of
LMXBs form with donor masses in the range $1-1.5\,$\,M$_\odot$, where the
differences between MB laws are more significant (see Fig.\,\ref{MBlaws}). 
As a result, the cumulative distributions $r(f>f_{\rm 0})$ for all the MB
laws (except for the extreme MB3) at high common--envelope efficiencies and
intermediate to high kick magnitudes are essentially indistinguishable
(Fig.\,\ref{r2}).

\section{COMPARISON WITH OBSERVATIONS}

In what follows, we constrain the three input parameters of our models
for LMXB formation by comparing the predicted values for the fraction of
short-period systems and the fraction of transients among them with
observations.  In order to do so, we estimate the {\em actual}
short-period fraction of LMXBs and the transient fraction among these,
based on the observed sample of LMXBs and the selection effects that
operate on them.

\subsection{Properties Derived from the Observed Sample}

In selecting the set of systems of interest to our study, we use the
catalogues compiled by Ritter \& Kolb (1998) and
van~Paradijs\markcite{vP95} (1995).  From the sample of {\em all}
observed LMXBs we exclude systems that are {\em not} found in the
Galactic disk and systems for which there has been evidence, based on
mass-function measurements, that the compact object is a black hole
(Tanaka \& Shibazaki\markcite{T96} 1996). Following van~Paradijs \&
White\markcite{vPW95} (1995) we also exclude non-variable sources with
X-ray fluxes lower than 10$\mu$Jy, because their nature is uncertain.
Our sample consists of (i) 39 systems that are classified as confirmed
neutron--star LMXBs because they show X-ray pulsations or type I bursts,
or are Z or atoll sources (Hasinger \& van~der~Klis\markcite{H89} 1989
for a definition), and (ii) 40 systems for which the nature of the 
compact object is uncertain and are considered as possible neutron stars. 

We expect that most of the persistently bright systems in the Galaxy have
already been detected; with a typical luminosity of $\simeq
10^{36}-10^{38}$\,erg\,s$^{-1}$, these sources are detectable with current
X-ray instruments, even if they are located at the far side of the Galaxy.
The situation is very different for the transient sources, for which the
quiescent luminosities are very low and the bright outburst state is
short-lived. In our sample, 31 systems appear to show  
transient behavior. 
Any estimate of the total number of transient neutron--star LMXBs
in the Galactic disk is clearly very uncertain.  We follow Tanaka \&
Shibazaki (1996) and assume that the number of transients is roughly equal
to the product of their average detection rate and their mean recurrence
time, $t_{\rm rec}$, further weighted by the completeness, $c$, of the
transient sample relative to the persistent sources. Assuming complete sky
coverage over the last 30 years (see Chen, Shrader, \& Livio\markcite{C97}
1997), we estimate the detection rate to be simply equal to the number of
transient systems detected over a thirty-year period divided by 30 years. 
The {\it observed} mean recurrence time of {\it all individual outbursts}
is $\simeq 2$\,yr (Chen et al.\ 1997).  This mean value is certainly biased
in favor of short recurrence times, as sources with only one recorded
outburst, hence long recurrence times, are not included and systems with
short recurrence time are overrepresented. Given the uncertainties, we
treat $t_{\rm rec}$ as a free parameter and obtain results for three
different values of it (1, 10, and 100\,yr), although we expect $t_{\rm
rec}$ to be longer than a few tens of years. We also treat the completeness
of the transient sample relative to the persistent sources as a free
parameter and investigate two cases, $c=0.5$ and $c=0.1$.  In fact, only
this relative completeness enters the observational estimate of the
fraction of short-period systems and that of transients among them, $s$ and
$r$ respectively, and not the assumption of 100\% completeness for
persistent sources. 

In our sample, the orbital period has been measured for 33 of the 79
systems, 24 of which are confirmed neutron--star LMXBs. This set of LMXBs
with measured periods is certainly biased by selection effects. Measurement
of orbital periods of LMXBs almost always relies on the identification of
the optical counterpart to the X-ray source and this is easier when the
donor star is bright, i.e., has evolved to the giant branch. Therefore, we
expect the majority of systems with undetermined period to have
main-sequence donors and hence short orbital periods. To investigate the
quantitative effect of this assumption we consider two extreme cases: (i)
all of them being short-period and (ii) all of them being long-period
systems. 

As in \S\,3, we use the orbital period, $P$, to distinguish between
j--driven ($P < P_{\rm crit}$) and n--driven ($P > P_{\rm crit}$) systems.
LMXBs with donors less massive than $\simeq 1$\,M$_\odot$ do not approach
the end of their main-sequence phase within the age of the Galaxy, have
periods below 12\,h, and are j--driven. Systems with masses in the range
$\simeq 1-1.5$\,M$_\odot$ (for 1.4\,M$_\odot$ neutron stars) and periods
longer than $\simeq 30$\,h have donors that have evolved beyond the base of
the giant branch and are n--driven (see Kalogera \& Webbink 1996). In the
intermediate regime, 12\,h\, $\lesssim P \lesssim $\,30\,h, both n--driven
and j--driven systems can occur. As a working criterion, we use $P_{\rm
crit} = 20$~h, although it turns out that our results are not sensitive to
this choice. 

In order to use the properties of the observed sample to evaluate the
models, we convert the observed numbers of systems into formation rates. For
this, we have to take into account the possible difference between the
lifetimes of systems in each group. Based on estimates of the lifetime of
LMXBs with donors on the giant branch (following Webbink, Rappaport, \&
Savonije\markcite{W83} 1983), we expect that n--driven and persistent
j--driven systems have comparable lifetimes.  Given the fact that transient
j--driven systems have mean transfer rates lower than some critical value,
and hence lower than those of persistent sources, their lifetimes are
correspondingly longer by an uncertain factor of a few. We again treat this
factor, $t_l$, as a free parameter, and consider the cases $t_l=1$ and
$10$. 

Table 1 summarizes the number of observed systems in each of four
different groups: transient and persistent sources among all LMXBs with
confirmed or possible neutron--star accretor (henceforth referred to as
``all LMXBs''), and also among only those with a confirmed neutron--star
accretor.  Based on the above discussion of selection effects acting on
the observed sample, we attempt to correct the numbers given in Table 1.
In Table 2, we list the fractions $r$ and $s$, calculated for the
observed sample that includes all LMXBs, for the different combinations
of $c$, $t_{\rm rec}$, and $t_l$.  The fractions are given under the
assumption that known LMXBs with undetermined orbital period are
short--period systems (columns 2 and 4 for $r$ and $s$, respectively) or
that they all are long--period systems (columns 3 and 5 for $r$ and $s$,
respectively). The corresponding numbers for the subset of LMXBs with
confirmed neutron--star primaries remain essentially unchanged.
Furthermore, we find that the inferred values of $r$ and $s$ are fairly
insensitive (typical changes of $\sim 10$\%) to uncertainties due to
Poisson noise in the observed number of systems in each group (Table 1). 

Based on what we consider as a reasonable choice of parameters, i.e.,
$t_{\rm rec} > 10$~yr, $c \lesssim 0.5$, $t_l\simeq10$, and with the
assumption that most LMXBs with unknown orbital period are short-period
systems, we adopt the values $r\approx 0.2$ (at $f>f_{\rm 0}=0.8$, see
section 2) and $s\approx 0.5$ as probable minimum requirements for the
transient and the short-period fractions, respectively. 

\subsection{Evaluation of the Models}

Here, we discuss the consistency of the models with respect to the limits
set by the observations, for three groups of the average kick magnitude
(see Fig.\,\ref{r1}, \ref{r2}, and \ref{s}). 

\paragraph{Negligible kicks} $\langle V_k^2 \rangle ^{1/2} \lesssim
50$\,km\,s$^{-1}$. All of the models result in very low values of $s$, with
only one exception, the case of strong MB (MB1a) and low $\ace$ (0.3).  For
this model though, magnetic braking is strong enough to prevent the
formation of a significant fraction of evolved MS donors and fails the
condition $r(f>0.8) > 0.2$. We have further investigated the dependence of
the MB1a models on $\ace$ and found that both of the constraints on $s$ and
$r$ can be satisfied {\em only} if $\ace$ is narrowly restricted at $\simeq
0.4$.  We note that the magnetic braking strength assumed by the MB1a law
at high stellar masses is not consistent with the rotational velocity data
of F stars, if the $\omega^3$ dependendence is valid for all $\omega$. 

\paragraph{Large kicks} $\langle V_k^2 \rangle ^{1/2} \ga
300$\,km\,s$^{-1}$. In this case {\em all} models can satisfy the
constraint for a high $s$ fraction; only the extreme MB3 (no magnetic
braking) models have some difficulty but cannot be excluded.  On the other
hand, MB3 models are the only ones that satisfy the constraint for
$r(f>0.8)>0.2$. All other MB models at $\ace\simeq 1$ can be safely
excluded, while at lower $\ace$ values the weak MB laws (MB1b, 2a, and 2b)
could be accepted only if the critical $f_{\rm 0}$ is smaller than $0.8$. 

\paragraph{Intermediate kicks} $\langle V_k^2 \rangle ^{1/2} \simeq
100-200$\,km\,s$^{-1}$. All models with high $\ace$ (close to unity) and
models without any magnetic braking (MB3) at any $\ace$ can be excluded
because they do not produce enough short-period systems. From the rest of
the models, only those with relatively weak magnetic braking (MB1b, 2a, and
2b) can produce enough j--driven systems with donors close to the end of
the main sequence, and hence are acceptable. 

Taking all the above into account, a globally consistent picture between
models and observations emerges for an average kick magnitude of about
$100-200$\,km\,s$^{-1}$, weak magnetic braking (MB1b, 2a, or 2b), and low
common--envelope efficiencies ($\lesssim 0.5-0.4$).  These models are the
only ones that are comfortably in agreement with observational constraints,
without the need to fine--tune model parameters.

\section{SUMMARY AND DISCUSSION}

We applied the population synthesis techniques developed by KW98 to
investigate the influence of varying mean supernova--kick magnitudes,
orbital angular momentum loss strengths, and common envelope efficiencies on
the population of transient and persistent short--period LMXBs that form
via the helium star supernova channel. A main premise of our study is the
identification of short--period transients with systems where the donor is
significantly nuclear--evolved, close to the end of core--hydrogen burning. 
We tested our models against two properties inferred from the observed
sample: that a significant fraction ($s \ga 50\%$) of nascent LMXBs are
short--period, and that a significant fraction of these ($r \ga 20\%$) have
donors close to the end of core--hydrogen burning.

We found that any combination of model parameters that results in a large
fraction of short-period systems with evolved donors, hence high values of
$r$, at the same time results in the formation of many systems with donors
that have evolved beyond the main sequence, and hence leads to low values of
$s$. It is exactly these two counteracting effects that allow us to
constrain the three model parameters.  With an increasing mean kick
magnitude $s$ grows, while $r$ drops. Model predictions for $s$ and $r$ are
consistent with observational estimates in an intermediate regime of
moderate mean kick magnitudes, $\vmean \simeq 100-200$~km/s, if (i)~the
orbital braking for systems with donor masses $1\la M_d \la 1.5\,\msun$ is
weak, i.e., much less effective than a simple extrapolation of magnetic
braking beyond $1\,\msun$ would suggest, and (ii)~the efficiency of common
envelope ejection is low ($\alpha_{\rm CE} \la 0.5$). 

Consistency with observational estimates could also be achieved in the
absence of kicks or for a very large mean kick velocity, but {\em only}
in combination with a fine-tuned common-envelope efficiency and/or
fairly extreme assumptions on the efficiency of magnetic braking at
masses $\ga 1\,\msun$ (very high for small kicks, almost negligible for
large kicks).  However, in view of the various uncertainties that enter
our study we cannot unambiguously rule out these models.  Overall, we
have shown that the model predictions are sufficiently sensitive to the
variation of the three input parameters to place constraints on them
when comparing to observations.  Given accurate input from the
observational sample, the constraints can be reasonably tight (for
example, had the adopted values of $s$ and $r$ been known within a
factor of 2, we could unambiguously exclude a single peak of kick
magnitudes at $\lesssim 50$ or $\gtrsim 300$\,km\,s$^{-1}$).
Nevertheless, significant uncertainties exist that have their origin
mainly in two areas:  secular evolution of j--driven LMXBs and
observational selection effects, which prevent us from accurately
inferring the intrinsic properties of the LMXB population from the
observed sample. 

In the first area, a systematic study of j--driven LMXB evolution with
detailed stellar models is needed to relate transient behavior
quantitatively to the degree of evolution of the donor for different
orbital angular momentum loss rates. This will also allow one to examine in
detail the transition regime between j--driven and n--driven systems and to
quantify the relative lifetimes of persistent and transient systems. 

The efficiency of magnetic braking affects both the lifetime of the systems
and the critical degree of evolution for transient behavior.  An increasing
braking strength restricts transient behavior to donors closer to the end
of core--hydrogen burning. At the same time, it leads to higher mass
transfer rates in persistent systems, and hence shorter relative lifetimes. 
The first effect increases the critical degree of evolution, $f_0$, whereas
the second reduces the intrinsic transient fraction, $r_0$, derived from
the observed sample. Therefore, the consistency criterion, $\rf > r_0$,
moves along a typical curve $\rf$ (Fig.\,\ref{r1} and \ref{r2}), so that
the choice of successful or unsuccessful models is not, to zeroth order,
affected by the neglect of these dependences. We note that another
long--standing problem in the evolution of LMXBs, the fate of j--driven
systems once they approach to orbital periods of 3\,h, and the apparent
lack of systems with $P \la 3$\,h (see KK97 for a discussion), does not
affect our study as long as any comparison is restricted to systems with
periods longer than 3\,h. 

The Skumanich--type magnetic braking formulation adopted here
is by no means the only possible parametrization of the orbital
angular momentum losses. Constraints on the real functional form could
come from further analysis of stellar rotational rates in open
clusters of different age. Some studies (e.g., Krishnamurthi et
al.\markcite{K97} 1997, and references therein)
suggest that $\dot J$ becomes a less steep function of
$\omega$ above a certain critical angular velocity. Repeating 
our model calculations with such more general magnetic braking laws seems
worthwhile only once the critical degree of evolution, $f_{\rm 0}$, for 
transient behavior can be estimated more quantitatively.

Progress in the second area of uncertainty requires a systematic study of
observational selection effects on X--ray binaries (given the X-ray
instruments used so far).  This is hampered by the difficulty of
calculating the spectral distribution of the emergent X--ray flux in a
system with given period and transfer rate.  The assessment of the relative
completeness of the different subgroups involved (black--hole,
neutron--star, transient, persistent, short--period, and long--period
systems) will certainly gain reliability when future observations increase
the known sample and the number of systems with determined binary and
transient parameters.  The completeness is further affected by a possible
dependence of the transient outburst recurrence time on orbital period.
Giant donor systems have larger disks, which take longer to reestablish the
critical pre--outburst surface density. If the recurrence time is
systematically longer in long--period systems, the short--period fraction
could be significantly smaller than estimated. This in turn would make
models with even smaller mean kick velocities acceptable. 

Despite these uncertainties, the preference for moderate mean kick
velocities of $\simeq 100-200$~km/s inferred from our study is likely to
persist even for a more detailed treatment of secular evolution and a
better understanding of selection effects.  This preference seems to be in
conflict with the fairly large {\em mean} natal kicks traditionally deduced
from pulsar proper motions (e.g., $\vmean \simeq 500$~km/s is favored by
Lorimer et al.\ 1997). (Note that our study cannot constrain the fraction
of very high ($\gtrsim 500$\,km\,s$^{-1}$) kick magnitudes, because this
only affects the absolute LMXB birth-rate normalization.) While this might
point to an underlying physical difference in the way supernova explosions
of type II and type Ib proceed, there is also the possibility that natal
pulsar velocities are significantly smaller than this estimate. Indeed, it
has been pointed out that a relatively wide variety of qualitatively
different distributions are consistent with the observed pulsar velocity
distribution (Hansen \& Phinney 1997; Cordes \& Chernoff\markcite{C98}
1998; Fryer et al.\ 1998; Hartmann\markcite{H98} 1998). Although our study
was limited to Maxwellian kick distributions we can conclude quite
generally that the observed short-period transient LMXB population
favors kick distributions with a dominant component at moderate mean
velocities of about 100 km/s.

We conclude by noting that with future progress the observational sample
of LMXBs will undoubtedly increase and improve in quality, and more
detailed calculations of the secular evolution of LMXBs will become
available. The model calculations we have presented here, of the
fraction of short-period LMXBs and the transient fraction among them,
can then be used to derive even tighter constraints on the supernova and
evolutionary parameters.

\acknowledgements 

It is a pleasure to thank Dimitrios Psaltis for many useful discussions
in the course of this study, as well as for a critical reading of the
manuscript.  We also thank the referee, Cole Miller, for several helpful
comments and suggestions. VK acknowledges partial support by the
Graduate College of the University of Illinois through a Dissertation
Completion Fellowship and by the Smithsonian Astrophysical Observatory
through a Harvard-Smithsonian Center for Astrophysics Postdoctoral
Fellowship. VK also is thankful for the hospitality of the Astronomy
Group at the University of Leicester.  ARK acknowledges support as a
PPARC Senior Fellow. Theoretical astrophysics research at Leicester is
supported by a PPARC rolling grant.

\clearpage

\begin{figure}[h]
\centerline{
\psfig{file=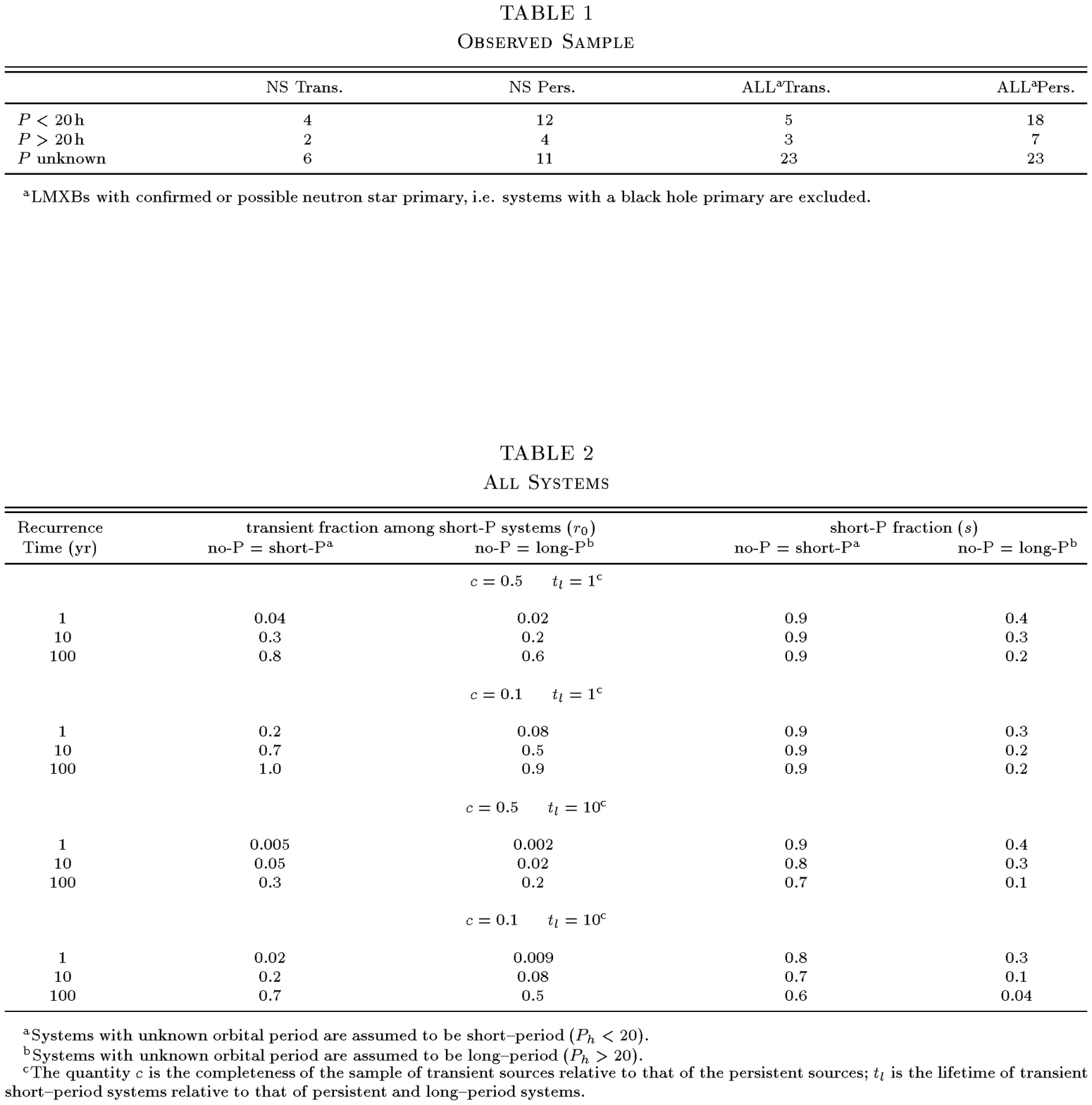,height=16truecm,width=15truecm}}
\end{figure}

\newpage

\begin{figure}[h]
\centerline{
\psfig{file=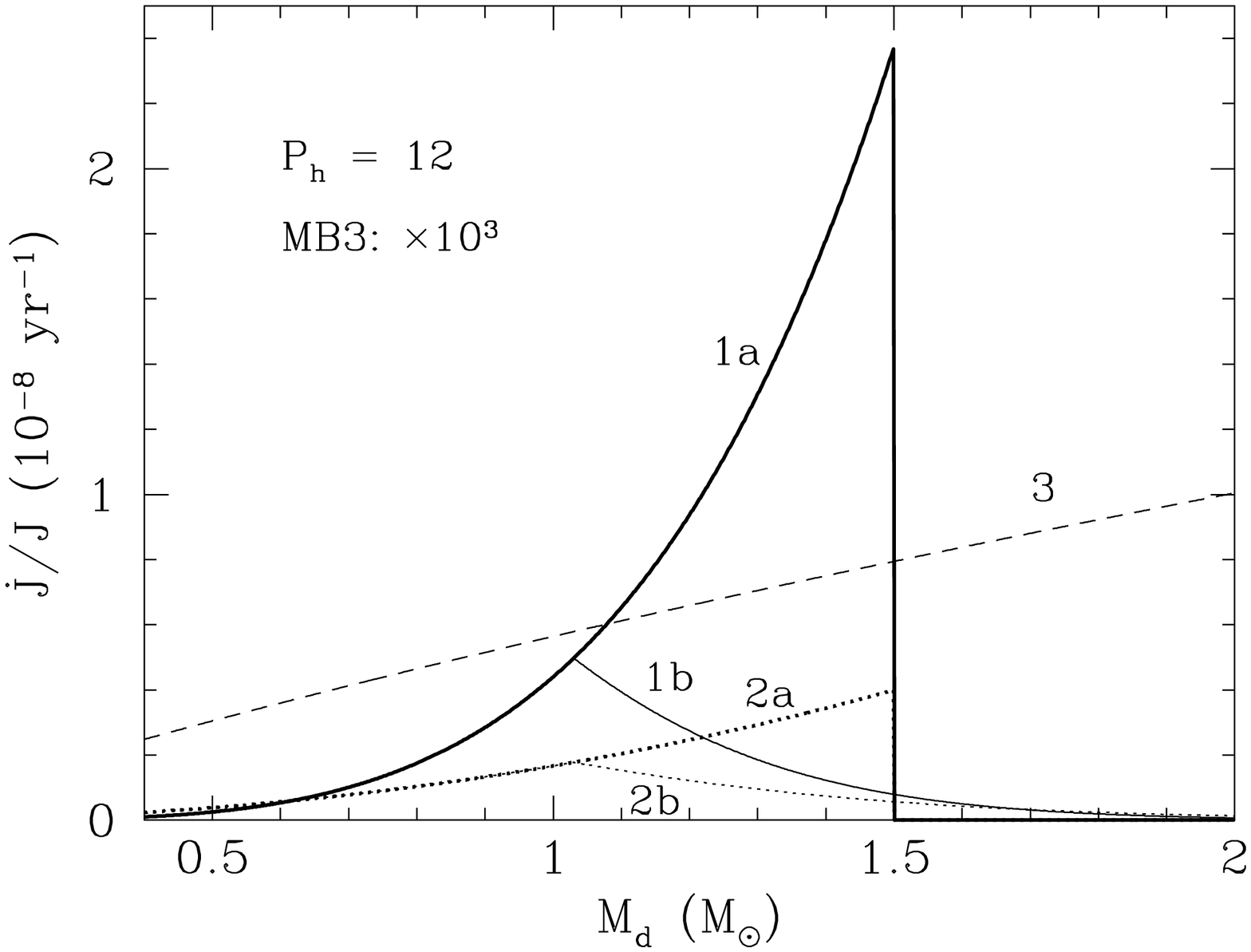,angle=-90,height=13truecm,width=16.5truecm}}
\end{figure}
\figcaption[] {Fractional orbital angular momentum loss rate, $\dot{J}/J$, as
a function of donor mass, $M_d$, for the five magnetic braking (MB)
laws tested in this study (neutron star mass $1.4\msun$, orbital
period $P_h$=12). The fractional rate for MB3 has been scaled by a
factor of $10^3$, for clarity.
The MB laws can be ordered according to decreasing strength (1a, 1b,
2a, 2b, 3) in the crucial regime $M_d\ga1\msun$.
\label{MBlaws}}

\newpage

\begin{figure}[h]
\centerline{
\psfig{file=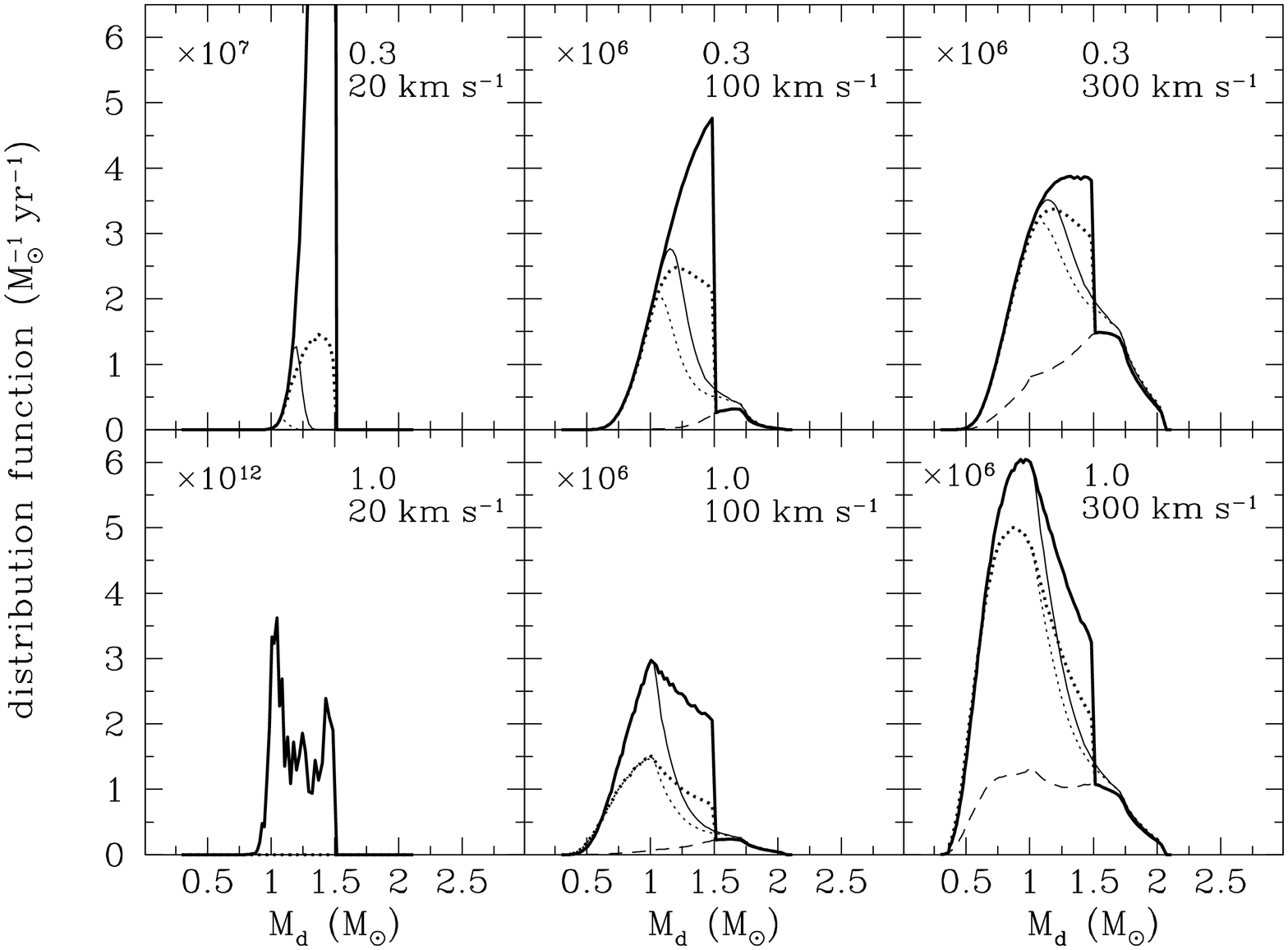,angle=-90,height=13truecm,width=17truecm}}
\end{figure}
\figcaption[] {Distribution over donor mass, $M_d$, of
the Galactic formation rate of LMXBs
for two values of the common-envelope efficiency (0.3 and
1.0) and three values of the r.m.s. kick magnitude (20, 100, and
300\,km\,s$^{-1}$), and for the five magnetic braking laws tested
in this study, MB1a ({\em thick solid line}), MB1b ({\em thin solid
line}), MB2a ({\em thick dotted line}), MB2b ({\em thin dotted line}), and
MB3 ({\em thin dashed line}). The curves in each panel have been scaled by
an appropriate factor for clarity. The case of $\alpha_{CE}=1.0$ and
$\langle V_k^2\rangle ^{1/2}=20\,$km\,s$^{-1}$, corresponds to LMXB models
with extremely low birth rates, which result in numerical inaccuracies in
the integration of the multi--dimensional distribution function.
For non--negligible kicks, the mean donor mass is smaller for large
$\alpha_{CE}$, and decreases with increasing kick magnitude. Only a few
systems have $M_d > 1.5\msun$.
\label{mdist}}

\newpage

\begin{figure}[h]
\centerline{
\psfig{file=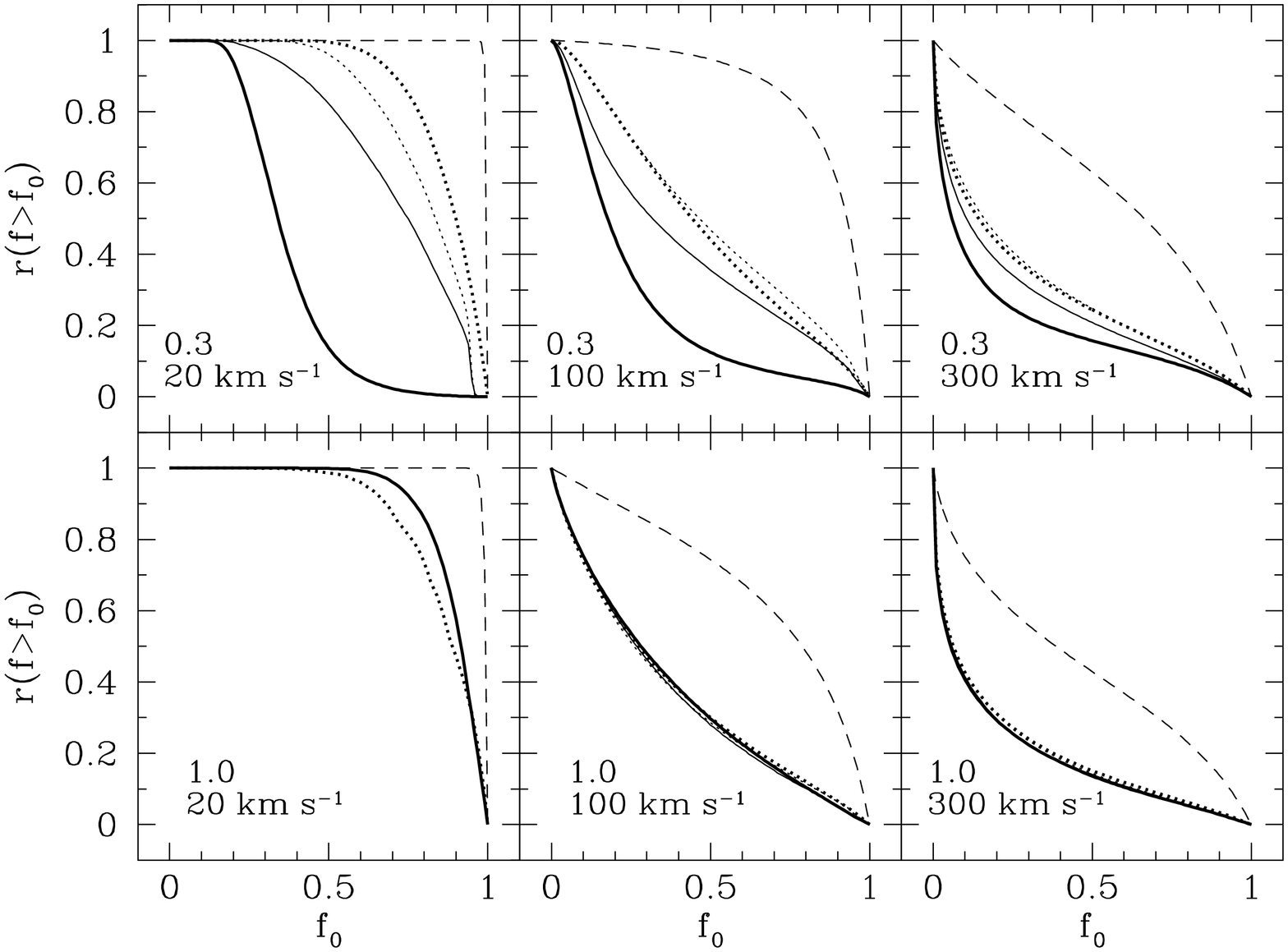,angle=-90,height=13truecm,width=17truecm}}
\end{figure}
\figcaption[] {Fraction, $\rf$, of j--driven LMXBs with donors with a degree
of evolution $f > f_0$, as a function of $f_0$, for two values of the
common-envelope efficiency (0.3 and 1.0) and three values of the r.m.s.
kick magnitude (20, 100, and 300\,km\,s$^{-1}$), and for the five magnetic
braking laws tested in this study. Line--type coding as in Figure 3. For the
case of $\alpha_{CE}=1.0$ and $\langle V_k^2\rangle
^{1/2}=20\,$km\,s$^{-1}$, the curves for MB1b and MB2b correspond to LMXB
models with unreasonably low birth rates and are not shown.
A system is assumed to be transient if $f$ is larger than a critical
value, probably $\ga 0.8$. Hence $\rf$ gives the transient fraction among
short--period LMXBs if $f_0$ is this critical value.
For non--negligible kicks, $\rf$ is insensitive to the kick magnitude
if $\alpha_{CE}=1.0$ (except for the extreme weak braking law MB3).
If $\alpha_{CE}$ is small, weaker MB laws give a larger fraction of
systems with large $f$. The case with negligible kicks is degenerate,
see text.  
\label{r1}}

\newpage

\begin{figure}[h]
\centerline{
\psfig{file=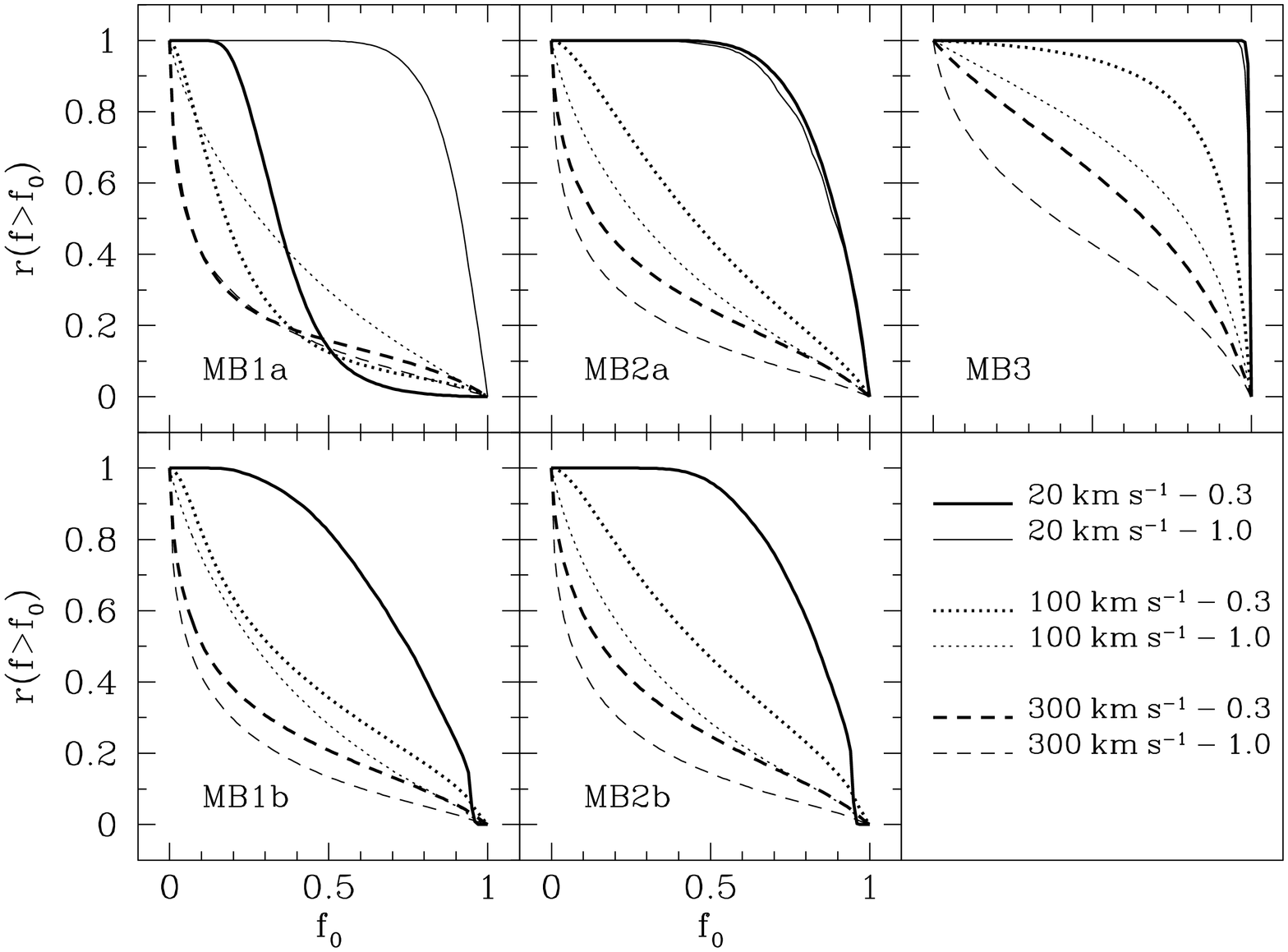,angle=-90,height=13truecm,width=17truecm}}
\end{figure}
\figcaption[] {Fraction, $\rf$, of j--driven LMXBs with donors with a degree
of evolution $f > f_0$, as a function of $f_0$, for the five magnetic
braking laws tested in this study, and for different values of the
common-envelope efficiency, 0.3 ({\em thick lines}) and 1.0 ({\em thin
lines}), and of the r.m.s. kick magnitude, 20\,km\,s$^{-1}$ ({\em solid
lines}), 100\,km\,s$^{-1}$ ({\em dotted lines}), and 300\,km\,s$^{-1}$
({\em dashed lines}). For the cases of MB1b and MB2b, the curves for
$\alpha_{CE}=1.0$ and $\langle V_k^2\rangle ^{1/2}=20\,$km\,s$^{-1}$
correspond to LMXB models with unreasonably low birth rates and are not
shown.
The fraction of systems with large $f$ (i.e.\ the transient fraction)
decreases with increasing kick magnitude, and is larger for small
$\alpha_{CE}$ (except for the very strong MB law 1a).
\label{r2}}

\newpage

\begin{figure}[h]
\centerline{
\psfig{file=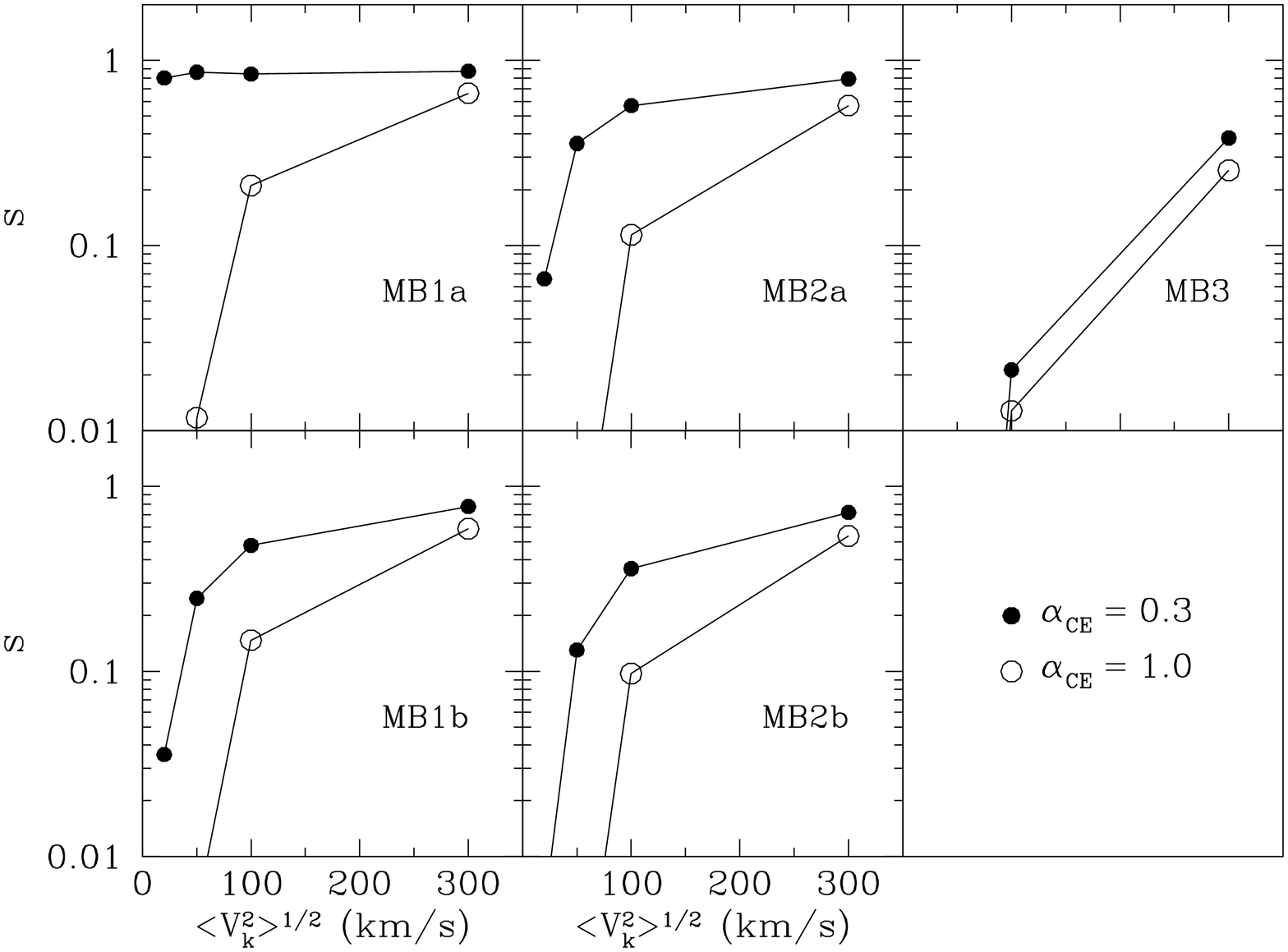,angle=-90,height=13truecm,width=17truecm}}
\end{figure}
\figcaption[] {Fraction, $s$, of short-period LMXBs as a function of the
r.m.s. kick magnitude, $\langle V_k^2\rangle ^{1/2}$, for the five magnetic
braking laws tested in this study, and for $\alpha_{CE}=0.3$ ({\em filled
circles}) and $\alpha_{CE}=1.0$ ({\em open circles}).
The fraction
$s$ increases with kick magnitude and is larger for models
with smaller $\alpha_{CE}$. When kicks are negligible $s$ is always
unrealistically small, except for the very strong MB law 1a.
\label{s}}

\end{document}